\documentclass[aps,prl,twocolumn,floatfix,superscriptaddress,showpacs]{revtex4-1}
\usepackage{graphicx,bm}
\usepackage[colorlinks,citecolor=blue]{hyperref}
\begin{document}

\title{Neutrino Nucleosynthesis of radioactive nuclei in supernovae}

\author{A. Sieverding}
\affiliation{Institut f\"ur Kernphysik (Theoriezentrum), Technische
  Universit\"at Darmstadt, Schlossgartenstra{\ss}e 2, 64289 Darmstadt,
  Germany} 
        
\author{L. Huther}
\affiliation{Institut f\"ur Kernphysik (Theoriezentrum), Technische
  Universit\"at Darmstadt, Schlossgartenstra{\ss}e 2, 64289 Darmstadt,
  Germany} 

\author{K.~Langanke}
\affiliation{Gesellschaft f\"ur Schwerionenforschung Darmstadt,
  Planckstr.~1, D-64259 Darmstadt, Germany}
\affiliation{Institut f\"ur Kernphysik (Theoriezentrum), Technische
  Universit\"at Darmstadt, Schlossgartenstra{\ss}e 2, 64289 Darmstadt,
  Germany} 

\author{G. Mart\'inez-Pinedo} 
\affiliation{Institut f\"ur Kernphysik (Theoriezentrum), Technische
  Universit\"at Darmstadt, Schlossgartenstra{\ss}e 2, 64289 Darmstadt,
  Germany} 
\affiliation{Gesellschaft f\"ur Schwerionenforschung Darmstadt,
  Planckstr.~1, D-64259 Darmstadt, Germany}

\author{A. Heger} 
\affiliation{Monash Centre for Astrophysics, School
  of Physics and Astronomy, Monash University, Victoria 3800,
  Australia}
\affiliation{University of Minnesota, School of Physics and Astronomy, Minneapolis, MN 55455, USA}
\affiliation{Shanghai Jiao-Tong University, CNA, Department of Physics and Astronomy, Shanghai 200240, P.R.~China}
\affiliation{The Joint Institute for Nuclear Astrophysics
}

\begin{abstract}
  We study the neutrino-induced production of nuclides in explosive
  supernova nucleosynthesis for progenitor stars with solar
  metallicity and initial main sequence masses between 15~M$_\odot$
  and 40~M$_\odot$.  We improve previous investigations i) by
  using a global set of partial differential cross sections for
  neutrino-induced charged- and neutral-current reactions on nuclei
  with charge numbers $Z < 76 $ and ii) by considering modern
  supernova neutrino spectra which have substantially lower average
  energies compared to those previously adopted in neutrino
  nucleosynthesis studies.  We confirm the production of $^7$Li,
  $^{11}$B, $^{138}$La, and $^{180}$Ta by neutrino nucleosynthesis,
  albeit at slightly smaller abundances due to the changed neutrino
  spectra. We find that for stars with a mass smaller than
  20~M$_\odot$, $^{19}$F is produced mainly by explosive
  nucleosynthesis while for higher mass stars it is produced by the
  $\nu$ process.  We also find that neutrino-induced reactions, either
  directly or indirectly by providing an enhanced abundance of light
  particles, noticeably contribute to the production of the
  radioactive nuclides $^{22}$Na and $^{26}$Al. Both nuclei are prime
  candidates for gamma-ray astronomy.  Other prime targets,
  $^{44}$Ti and $^{60}$Fe, however, are insignificantly produced by
  neutrino-induced reactions.  We also find a large increase in the
  production of the long-lived nuclei $^{92}$Nb and $^{98}$Tc due to
  charged-current neutrino capture.
\end{abstract}


\maketitle

Astrophysical objects like stars, novae, or supernovae are the origin
of most of the elements in the
Universe~\cite{Burbidge.Burbidge.ea:1957,Cameron:1957}. Whereas the
likely nucleosynthesis processes associated with these objects have
been identified and a general understanding has been developed, many
details of their operation are still
unresolved~\cite{Wiescher.Kaeppeler.Langanke:2012,Thielemann.Arcones.ea:2011,Arnould.Takahashi:1999}.
This is due to limited computational capabilities to simulate
astrophysical objects and to the fact that the properties of the nuclides
involved in the nucleosynthesis processes are not known experimentally
and have to be modeled~\cite{Langanke.Schatz:2013}.

Detections of gamma-rays from radioactive nuclei by space bound
observatories like INTEGRAL~\cite{Winkler.Diehl.ea:2011} are an
invaluable tool to determine their production sites and thereby
advance our understanding of astrophysical nucleosynthesis.  Such
detection allows for a snapshot view of the ongoing nucleosynthesis in
our galaxy and, provided a suited nuclear half-life, to relate the
origin of the nuclide to a specific astrophysical
source~\cite{Diehl.Halloin.ea:2006}. In cases where the observation
can be assigned to a particular supernova remnant, one can learn about
asymmetries in the explosion~\cite{Grefenstette.Harrison.ea:2014}.  The
prime nuclide for gamma-ray astronomy in recent years has been
$^{26}$Al~\cite{Diehl:2013}. Its production has been associated with
several astrophysical sources (see
ref.~\cite{Woosley.Hartmann.ea:1990} and references therein),
however, in recent years evidence has been brought
forward~\cite{Timmes.Woosley.ea:1995,Diehl.Timmes:1998,Diehl:2013}
that massive stars can account for most of the $^{26}$Al in the
galaxy.  Other gamma-ray astronomy candidates like $^{22}$Na,
$^{44}$Ti, and $^{60}$Fe are also related to core-collapse supernovae
\cite{Iyudin.Diehl.ea:1994,Timmes.Woosley.ea:1995,Limongi.Chieffi:2006,Woosley.Heger.Weaver:2002,Rauscher.Heger.ea:2002}.

It has long been recognized that, in the $\nu$ process during a
supernova explosion, neutrino-nucleus reactions are essential for the
synthesis of selected nuclides like $^{7}$Li, $^{11}$B, $^{15}$N,
$^{19}$F, $^{138}$La, or
$^{180}$Ta~\cite{Woosley.Hartmann.ea:1990,Heger.Kolbe.ea:2005} or can
contribute to the production of long lived radioactive nuclides
\cite{Woosley.Hartmann.ea:1990,Timmes.Woosley.ea:1995,Woosley.Heger.Weaver:2002,Rauscher.Heger.ea:2002}.
In the $\nu$ process neutrinos of all flavors, which are emitted by
the cooling proto-neutron star (PNS), interact with nuclei as they
pass through the surrounding stellar matter.  At the same time, these
outer layers are heated up and compressed by the explosion shockwave
propagating outward from the PNS and causing the ejection of the
material.  Neutral-current reactions excite the nucleus to states
above particle thresholds so that the subsequent decay is accompanied
by emission of light particles (proton, neutron or $\alpha$
particle). Due to the relatively low energies of the neutrinos,
charged-current reactions can only be induced by electron-type
neutrinos. This process can be accompanied by light-particle emission
if the $(\nu_e,e^-)$ or $(\bar{\nu}_e, e^+)$ reactions excite the
daughter nucleus to levels above particle thresholds. Hence selected
nuclei, e.g. $^{11}$B, $^{19}$F, $^{138}$La, and $^{180}$Ta, are
produced directly as daughter products of neutrino-induced
reactions. The abundance of other nuclides, e.g. $^{7}$Li, is enhanced
indirectly by neutrino spallation reactions as these increase the
amount of light particles required to synthesize these nuclides within
a network of charged-particle reactions.

The focus of this letter is to explore the impact of the $\nu$ process
on the production of long-lived radioactive nuclei of interest to
gamma-ray astronomy.  Previous investigations of nucleosynthesis by
neutrino-induced reactions have been based on stellar simulations
using various hydrodynamical
models~\cite{Woosley.Hartmann.ea:1990,Heger.Kolbe.ea:2005,Limongi.Chieffi:2006}
using neutrino-nucleus cross section data which were restricted to a
set of key nuclei (like those which are quite abundant in outer
burning shells) and to a limited number of decay channels. Furthermore,
the simulations adopted supernova neutrino energy spectra, described
by Fermi-Dirac distributions with chemical potential $\mu=0$ and
temperature $T_\nu$, which were appropriate at the time the studies
were performed; i.e. $T_{\nu_e,\bar{\nu}_e}=4$-5~MeV for electron
(anti-)neutrinos, corresponding to average energies, $\langle
E_\nu\rangle = 3.15\, T_\nu$, between 12~MeV and
16~MeV~\cite{Woosley.Hartmann.ea:1990,Heger.Kolbe.ea:2005} and
$T_{\nu_{\mu,\tau}}=5$-10~MeV~\cite{Woosley.Hartmann.ea:1990,%
Timmes.Woosley.ea:1995,Heger.Kolbe.ea:2005} for muon and
tau neutrinos as well as for the corresponding anti-neutrinos,
corresponding to average energies between 16~MeV and 32~MeV.  We improve
these simulations in two relevant aspects.  Firstly, we have derived a
complete set of partial differential cross sections for
neutrino-induced charged- and neutral-current reactions on the global
chart of nuclei for charge numbers $Z < 76$ considering various
single- and multi-particle decay channels.  Secondly, the more
realistic treatment of neutrino transport in recent supernova
simulations~\cite{Huedepohl.Mueller.ea:2010,Martinez-Pinedo.Fischer.ea:2012,Martinez-Pinedo.Fischer.Huther:2014}
yield spectra for all neutrino families which are noticeably shifted
to lower energies. This reduces the neutrino-nucleus cross sections;
in particular particle spallation cross sections for neutral-current
reactions which are very sensitive to the tail of the neutrino
spectra. Our choice of neutrino temperatures is $T_{\nu_e}= 2.8$~MeV,
$T_{\bar{\nu}_e}=T_{\nu_{\mu,\tau}}= 4.0$~MeV based on recent
simulations~\cite{Huedepohl.Mueller.ea:2010,Martinez-Pinedo.Fischer.ea:2012,Martinez-Pinedo.Fischer.Huther:2014}.

We have calculated partial differential neutrino-nucleus cross
sections globally for nuclei with $Z < 76$ based on a two-step
strategy~\cite{Kolbe.Langanke.ea:1992}: i) the neutrino-induced
nuclear excitation cross sections to a final state at energy $E$ have
been calculated within the Random Phase Approximation (following
\cite{Kolbe.Langanke.ea:2003}) allowing for partial proton and neutron
occupancies and considering multipole transitions up to order $\lambda
=4$. The single particle energies were adopted from an appropriate
Woods-Saxon parametrization, adjusted to reproduce the proton and
neutron thresholds and to account for the energies of the Isobaric
Analog State and the leading giant resonances. ii) The decay
probabilities of the excited nuclear levels have been derived within
the statistical model. At low excitation energies we use a Modified
Smoker code~\cite{Loens:2010} which considers experimentally known
states and their properties explicitly and then matches the
experimental spectrum to a level density.  The code is restricted to
treat single-particle decays.  To allow for multi-particle decay,
which becomes relevant at modest excitation energies or in nuclei with
large neutron excess and hence small separation energies, we have
adopted the ABLA code~\cite{Kelic.Valentina.Schmidt:2009} at higher
excitation energies, which has been well validated to properly
describe multi-particle decays and fission. The results of the two
statistical model codes have been smoothly matched at moderate
energies above the single-particle thresholds. In the reaction network
all neutrino-induced reactions on nuclei with charge number $Z<76$ are
included.  This gives a consistent picture of $\nu$-nucleosynthesis
covering the whole range of nuclei from light to heavy.  Crucial
cross-sections for $^{4}$He are taken from
reference~\cite{Gazit.Barnea:2007} and for $^{12}$C the values used in
reference~\cite{Woosley.Hartmann.ea:1990} are adopted. Our cross
sections for $^{20}$Ne and $^{138}$Ba and $^{180}$Hf are consistent
with experimental
constraints~\cite{Heger.Kolbe.ea:2005,Byelikov.Adachi.ea:2007}.

The evolution of the shockwave passing through the outer layers of the
star is described using the parametrization of
ref.~\cite{Woosley.Hartmann.ea:1990}. It reproduces hydrodynamical
calculations~\cite{Woosley.Heger.Weaver:2002} particularly for the
peak temperature reached as the shock passes. This temperature is the
key quantity for nucleosynthesis. The parametrization assumes that the
region behind the shock is radiation dominated, containing the kinetic
explosion energy of $10^{51}$ erg.  The neutrino luminosity is modeled
following reference~\cite{Woosley.Hartmann.ea:1990}. It assumes a
total energy of $3\times 10^{53}$~ergs equally distributed in all
neutrino flavors. For a particular neutrino flavor the luminosity is
assumed to decay exponentially with a timescale of 3~s, i.e. $L_\nu =
5/3\times 10^{52}\exp(-t/3)$~erg~s$^{-1}$ with $t$ in seconds. The
composition is followed with a reaction network including all relevant
nuclei and reactions up charge number $Z=76$. The abundances are
evolved up to $2.5\times 10^4$~s after bounce. We have used
pre-supernova progenitor models from
ref.~\cite{Rauscher.Heger.ea:2002,Heger.2sn.org} in the mass range
15-40~M$_\odot$. It is unclear which of the explored models will
explode and how the explosion energy and amount of fallback depend on
progenitor mass and
structure~\cite{Woosley.Weaver:1995,Horiuchi.Nakamura.ea:2014,%
Sukhbold.Woosley:2014,Ertl.Janka.ea:2015}. We
find that the $\nu$ process mainly operates in outer regions of the
stellar mantle that should not be affected by fallback. Nevertheless,
fallback may trigger the formation of a black hole resulting in a
sudden end of neutrino emission~\cite{Fischer.Whitehouse.ea:2009} This
possibility is neglected in our calculations.

\begin{table}[Htb]
  \caption{Production factors relative to solar abundances from
    reference \cite{Lodders:2003}, normalized to $^{16}$O
    production. Shown are the results obtained without neutrino, with
    our choice of neutrino 
    temperatures (``Low energies''), and  with the choice of
    ref.~\cite{Heger.Kolbe.ea:2005} (``High energies'').\label{tab:prodfac}}  
  \begin{ruledtabular}
    \begin{tabular}{llccc}
      Star & Nucleus& no $\nu$ & Low energies\footnote{$T_{\nu_e}=
       2.8$~MeV, $T_{\bar{\nu}_e}=T_{\nu_{\mu,\tau}}= 4.0$~MeV}& High
     energies\footnote{$T_{\nu_e}=T_{\bar{\nu}_e}=4.0$~MeV,
       $T_{\nu_{\mu,\tau}}=6.0$~MeV}\\  \hline      
     15~M$_\odot$ & $^{7}$Li& 0.001 & 0.28 	& 2.54	\\ 
     & $^{11}$B&0.007 & 1.43  & 6.13 \\ 
     & $^{15}$N&0.67 & 0.68  & 0.79 \\ 
     & $^{19}$F& 1.02 & 1.14  & 1.31 \\ 
     & $^{138}$La &0.07 & 0.67  & 1.18 \\ 
     & $^{180}$Ta& 0.07 & 1.14 & 1.81 \\ \hline 
     25~M$_\odot$ & $^{7}$Li& 0.0005 & 0.11 	& 0.55	\\ 
     & $^{11}$B&0.003 & 0.80  & 2.61 \\ 
     & $^{15}$N&0.08 & 0.10  & 0.13 \\ 
     & $^{19}$F& 0.06 & 0.24  & 0.43 \\ 
     & $^{138}$La &0.03 & 0.63  & 1.14 \\ 
     & $^{180}$Ta& 0.14 & 1.80 & 2.81 \\ 
   \end{tabular}
 \end{ruledtabular}
\end{table}

The main candidates for neutrino nucleosynthesis are $^{7}$Li,
$^{11}$B, $^{15}$N, $^{19}$F, $^{138}$La, and $^{180}$Ta
\cite{Heger.Kolbe.ea:2005}, all of which are observed in the solar
system, but are not produced in sufficient amount by supernova
simulations without including neutrino interactions. Neutrino
nucleosynthesis pushes the production factors of those nuclei close to
the solar system values (Table~\ref{tab:prodfac}).  The relative
increase of $^{7}$Li and $^{11}$B is strongly affected by the $\nu$
process.  At the base of the He-shell the neutral current
neutrino-interactions $^{4}$He$(\nu,\nu'p)$ and $^{4}$He$(\nu,\nu'n)$
contribute to produce $^{7}$Li by the reactions
$^{3}$He($\alpha$,$\gamma$)$^{7}$Be($\beta^+$)$^{7}$Li and $^{11}$B
via $^{3}$H($\alpha$,$\gamma$)$^{7}$Li($\alpha$,$\gamma$)$^{11}$B.
The yields obtained in our calculations are consistent with
reference~\cite{Heger.Kolbe.ea:2005}, although our neutrino energies
are substantially lower.  We find that $^{11}$B can be produced in
full solar abundance, whereas $^{7}$Li is still underproduced,
supporting the need for other sources of
$^{7}$Li~\cite{Heger.Woosley:2010}.


\begin{table*}[Htb]
  \caption{Yields of different radioactive nuclei and the effects of
    the $\nu$-process for different progenitor stars. Shown are the
    results for the calculations without including neutrino
    interactions (``no $\nu$'') and the results including
    neutrinos. \label{tab:prodall}}  
  \begin{ruledtabular}    
    \begin{tabular}{lcccccccccc}
      &  \multicolumn{2}{c}{15 M$_\odot$} & \multicolumn{2}{c}{20
        M$_\odot$} &   \multicolumn{2}{c}{25 M$_\odot$}&
      \multicolumn{2}{c}{30 M$_\odot$}\\
      \cline{2-3}\cline{4-5}\cline{6-7}\cline{8-9}  
      nucleus &  no $\nu$&  $\nu$& no $\nu$ & $\nu$ & no $\nu$ & $\nu$ &
      no $\nu$ & $\nu$ \\ \hline 
      $^{6}$Li   &5.46$\times 10^{-11}$  & 5.65$\times 10^{-11}$ &
      4.24$\times 10^{-17}$  & 5.01$\times 10^{-12}$ & 8.78$\times
      10^{-11}$  & 9.87$\times 10^{-11}$  & 1.04$\times 10^{-10}$  &
      1.24$\times 10^{-10}$    \\ 
      $^{7}$Li   &1.44$\times 10^{-09}$  & 1.04$\times 10^{-07}$ &
      5.29$\times 10^{-14}$  & 1.84$\times 10^{-07}$ & 2.35$\times
      10^{-09}$  & 7.25$\times 10^{-08}$  & 2.74$\times 10^{-09}$  &
      8.89$\times 10^{-08}$    \\ 
      $^{9}$Be   &5.22$\times 10^{-11}$  & 7.09$\times 10^{-11}$ &
      2.73$\times 10^{-19}$  & 8.81$\times 10^{-12}$ & 8.01$\times
      10^{-11}$  & 1.15$\times 10^{-10}$  & 9.16$\times 10^{-11}$  &
      1.17$\times 10^{-10}$    \\ 
      $^{10}$B   &1.15$\times 10^{-09}$  & 2.06$\times 10^{-09}$ &
      1.60$\times 10^{-10}$  & 5.34$\times 10^{-10}$ & 2.09$\times
      10^{-09}$  & 3.84$\times 10^{-09}$  & 2.29$\times 10^{-09}$  &
      3.42$\times 10^{-09}$    \\ 
      $^{11}$B   &3.70$\times 10^{-09}$  & 1.10$\times 10^{-07}$ &
      6.74$\times 10^{-11}$  & 2.23$\times 10^{-07}$ & 6.33$\times
      10^{-09}$  & 1.97$\times 10^{-07}$  & 7.20$\times 10^{-09}$  &
      1.26$\times 10^{-07}$    \\ 
      $^{19}$F   &6.10$\times 10^{-05}$  & 6.43$\times 10^{-05}$ &
      6.94$\times 10^{-06}$  & 8.24$\times 10^{-06}$ & 1.40$\times
      10^{-05}$  & 5.53$\times 10^{-05}$  & 9.19$\times 10^{-05}$  &
      1.32$\times 10^{-04}$    \\ 
      $^{22}$Na  &2.48$\times 10^{-07}$  & 7.50$\times 10^{-07}$ &
      4.71$\times 10^{-07}$  & 8.13$\times 10^{-07}$ & 3.85$\times
      10^{-06}$  & 4.94$\times 10^{-06}$  & 7.97$\times 10^{-06}$  &
      8.71$\times 10^{-06}$    \\ 
      $^{26}$Al  &1.61$\times 10^{-05}$  & 2.43$\times 10^{-05}$ &
      1.89$\times 10^{-05}$  & 2.36$\times 10^{-05}$ & 5.10$\times
      10^{-05}$  & 7.11$\times 10^{-05}$  & 2.02$\times 10^{-05}$  &
      3.63$\times 10^{-05}$    \\ 
      $^{36}$Cl  &5.64$\times 10^{-07}$  & 2.22$\times 10^{-05}$ &
      1.11$\times 10^{-04}$  & 1.34$\times 10^{-04}$ & 4.78$\times
      10^{-06}$  & 5.23$\times 10^{-05}$  & 1.30$\times 10^{-06}$  &
      3.55$\times 10^{-05}$    \\ 
      $^{44}$Ti  &1.09$\times 10^{-04}$  & 1.23$\times 10^{-04}$ &
      5.38$\times 10^{-05}$  & 6.15$\times 10^{-05}$ & 1.13$\times
      10^{-04}$  & 1.09$\times 10^{-04}$  & 5.67$\times 10^{-05}$  &
      7.84$\times 10^{-05}$    \\ 
      $^{60}$Fe  &1.35$\times 10^{-04}$  & 1.40$\times 10^{-04}$ &
      3.55$\times 10^{-05}$  & 3.55$\times 10^{-05}$ & 1.60$\times
      10^{-04}$  & 1.50$\times 10^{-04}$  & 7.23$\times 10^{-05}$  &
      7.49$\times 10^{-05}$    \\ 
      $^{92}$Nb  &2.15$\times 10^{-12}$  & 4.09$\times 10^{-11}$ &
      6.59$\times 10^{-10}$  & 7.64$\times 10^{-10}$ & 9.54$\times
      10^{-11}$  & 6.05$\times 10^{-10}$  & 4.58$\times 10^{-12}$  &
      4.27$\times 10^{-10}$    \\ 
      $^{98}$Tc  &1.12$\times 10^{-11}$  & 1.37$\times 10^{-11}$ &
      1.11$\times 10^{-11}$  & 3.31$\times 10^{-11}$ & 1.11$\times
      10^{-11}$  & 5.89$\times 10^{-11}$  & 1.74$\times 10^{-12}$  &
      4.07$\times 10^{-11}$    \\	  
      $^{138}$La &1.34$\times 10^{-11}$  & 1.32$\times 10^{-10}$ &
      2.11$\times 10^{-10}$  & 3.07$\times 10^{-10}$ & 2.79$\times
      10^{-11}$  & 5.17$\times 10^{-10}$  & 2.97$\times 10^{-11}$  &
      4.15$\times 10^{-10}$    \\ 
      $^{180}$Ta &7.89$\times 10^{-14}$  & 1.85$\times 10^{-12}$ &
      1.49$\times 10^{-13}$  & 2.32$\times 10^{-12}$ & 8.09$\times
      10^{-13}$  & 1.23$\times 10^{-11}$  & 3.05$\times 10^{-13}$  &
      1.31$\times 10^{-11}$    \\ 
    \end{tabular}
  \end{ruledtabular}
\end{table*}

Reference \cite{Heger.Kolbe.ea:2005} states that the $\nu$ process can
probably not account for the entire solar abundance of $^{19}$F. We
find that the mechanism for its production is rather different for the
low mass and high mass progenitors considered here.  For the
15~M$_\odot$ model, the ratio $^{19}$F/$^{16}$O is consistent with
solar proportions even without neutrinos. The pre-supernova
$^{19}$F/$^{16}$O ratio of 0.07 is increased to a final value of
1.14 during shock passage mainly by the reaction sequence
$^{18}$O$(p,\alpha)^{15}$N$(\alpha,\gamma)^{19}$F operating on
$^{18}$O at the lower edge of the He-shell where post shock
temperatures reach values up to 0.7~GK at densities of up to
1500~g~cm$^{-3}$. The temperature and density reached in this region
depends on the radial position of the shell interface and hence is
very sensitive to the progenitor
structure. Reference~\cite{Rauscher.Heger.ea:2002} discusses the major
changes of stellar structure that appear for stars between 15 and
25~M$_\odot$. In particular, we find that for stars with a mass
smaller than 20~M$_\odot$ $^{19}$F is produced mainly by explosive
nucleosynthesis, whereas for higher mass stars it is produced by the
$\nu$ process (Table~\ref{tab:prodall}). Given the uncertainties
involved in stellar modeling, arising especially from the treatment of
convection and uncertainties in nuclear reaction rates at
astrophysical energies, the production of $^{19}$F up to solar
abundance cannot be excluded by our calculations. To address the
sensitivity to the progenitor structure, we have explored the
production of $^{19}$F based on models from
ref.~\cite{Limongi.Chieffi:2006,Limongi.Chieffi.url} and found similar
enhancement in the production factors of $^{19}$F for low mass stars
with substantial quantitative differences with respect to the yields
presented here. Our results for $^{138}$La and $^{180}$Ta are
consistent with those of
refs.~\cite{Heger.Kolbe.ea:2005,Byelikov.Adachi.ea:2007}.

\begin{figure}[htb]
 \includegraphics[width=\linewidth]{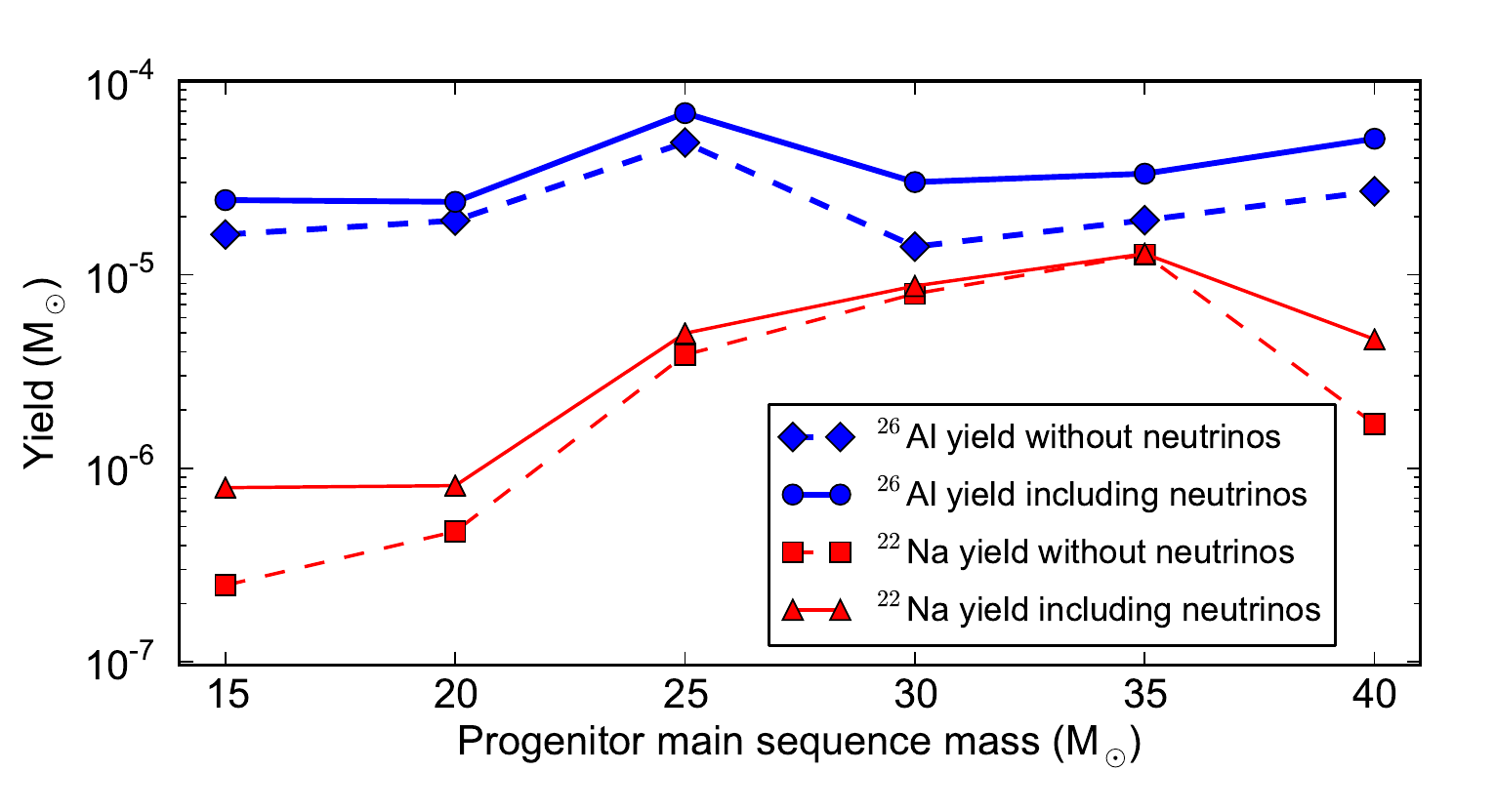}
 \caption{Yields of $^{26}$Al (thick blue lines with circles) and
   $^{22}$Na (thin red lines with triangles) for the set of progenitor
   stars considered. Given is the yield without neutrinos (dashed
   lines) and including neutrinos with $T_{\nu_e}=2.8$~MeV and
   $T_{\bar{\nu}_e}=T_{\nu_\mu,\nu_\tau}=4$~MeV.\label{fig:al26na22}}
\end{figure}

\begin{figure}[htb]
  \includegraphics[width=\linewidth]{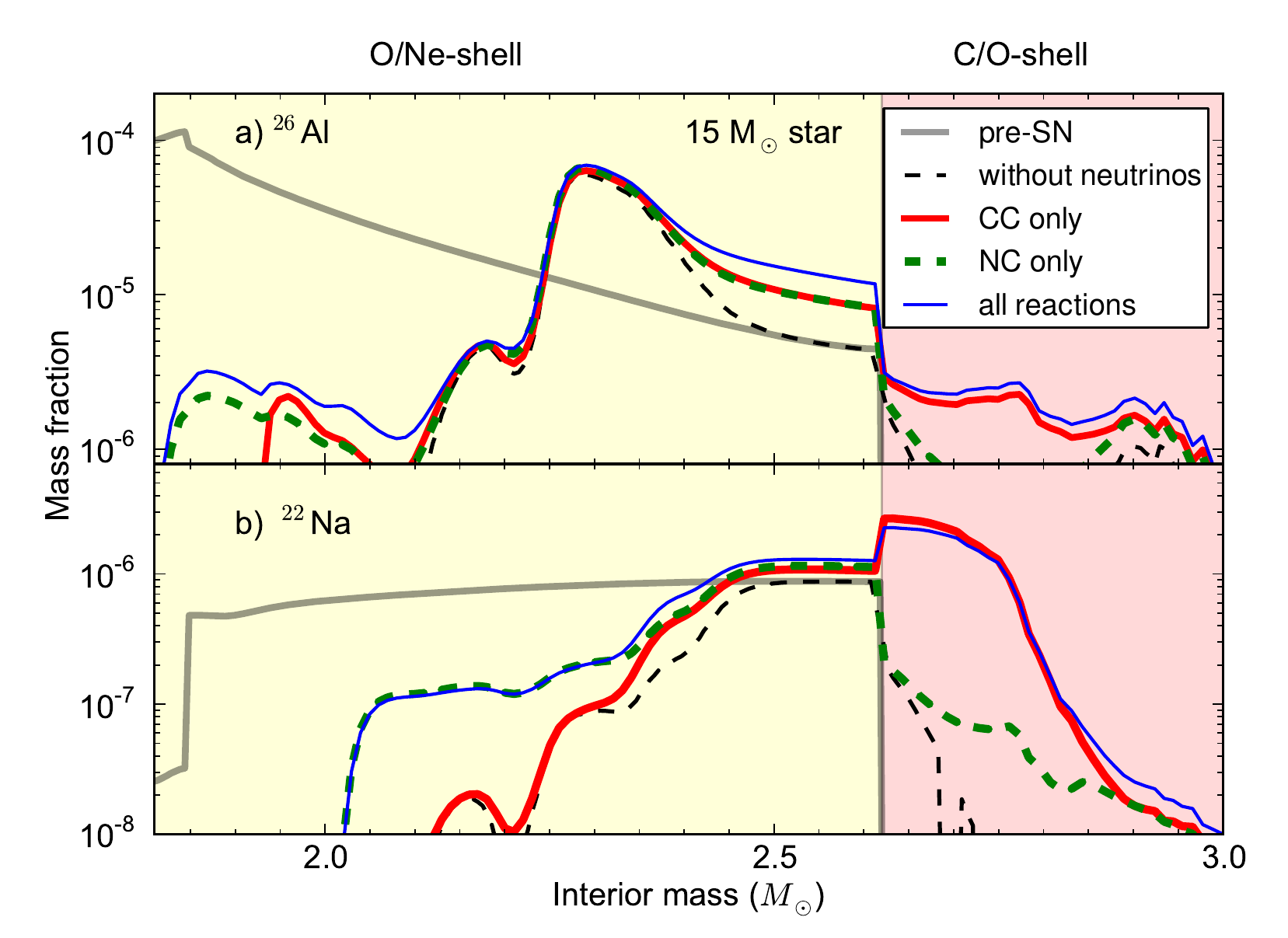}
  \caption{Mass fraction of $^{26}$Al (upper panel) and $^{22}$Na
    (lower panel) for the same 15 M$_\odot$ main sequence mass
    progenitor star of solar metallicity. Shown are the results for
    calculations with and without including neutrino interactions,
    with charged-current reactions only, and neutral-current reactions
    only. The pre-supernova mass fractions are also shown.\label{fig:al26s15}}
\end{figure}

Table~\ref{tab:prodall} lists the nucleosynthesis yields for those
nuclei that are affected by the $\nu$ process including long lived
radioactive nuclei. The yields of short-lived radioactive nuclei, e.g.
$^{32}$P, $^{72}$As, $^{84}$Rb, $^{88}$Y are increased by factors
between 10 and 100. Their lifetimes are of the order of a 100 days or
shorter, putting their decay signal in competition with $^{56}$Ni and
its daughter $^{56}$Co which by far dominates the early lightcurve and
therefore outshines the signature of the $\nu$ process. The typical
yields for $^{72}$As, $^{84}$Rb, and $^{88}$Y are $10^{-8}$~M$_\odot$, which
may allow for the observation of the gamma-ray decay lines.  We also
find a significant enhancement of the production of the long-lived
isotope $^{36}$Cl which, however, decays mainly to the ground-state
of $^{36}$Ar without characteristic $\gamma$-rays. The yield of
$^{26}$Al is known to be enhanced by neutrino
nucleosynthesis~\cite{Woosley.Hartmann.ea:1990,Timmes.Woosley.ea:1995}. We
find that the yield of $^{26}$Al is increased by factors between 1.25
and 2.51 in the range of progenitor models studied (see
Table~\ref{tab:prodall} and Figure~\ref{fig:al26na22}). The production
of $^{26}$Al during the explosion occurs mostly in a narrow region of
the O/Ne shell, in which $^{26}$Mg and $^{25}$Mg are abundant and the
post-shock temperature is below 2~GK. Deeper layers are subject to
higher peak temperatures such that the $^{26}$Al produced before the
explosion is destroyed by $^{26}$Al$(p,\gamma)$.  Neutrinos contribute
to the production of $^{26}$Al during the explosive phase by two
different mechanisms. Neutrino-induced spallation reactions on the
most abundant nuclei in the O/Ne shell, $^{20}$Ne, $^{24}$Mg, and
$^{16}$O increase the number of free protons, enhancing the reaction
$^{25}$Mg$(p,\gamma)$, which is also the main production channel 
without neutrinos. Additionally, the charged-current reaction
$^{26}$Mg$(\nu_e,e^-)$ gives significant
contributions. Figure~\ref{fig:al26s15} illustrates the different
production channels for the 15 M$_\odot$ progenitor model. Compared
with previous studies, we find a reduction of the neutral-current
channel due to the reduced neutrino energies. Hence, both
charged- and neutral-current reactions contribute to a similar extent
to the production of $^{26}$Al in the O/Ne layer. The enhancement of
the $^{25}$Mg$(p,\gamma)$ is confined to a narrow region of optimal
temperature, whereas the $^{26}$Mg$(\nu_e,e^-)$ contributes more evenly
throughout the entire layer, decreasing with the neutrino flux at
larger radii.


Our 20 M$_\odot$ progenitor suffers an early merging of the convective
O-, Ne-, and C-shells which significantly changes the chemical
composition of this model~\cite{Rauscher.Heger.ea:2002} and depletes
the progenitor abundances of $^{26}$Mg and $^{25}$Mg. Consequently,
the yield of $^{26}$Al is reduced. The 25~M$_\odot$ progenitor
exhibits the largest compactness, i.e., the mass over radius ratio for
a particular enclosed mass, and therefore provides the most favorable
conditions for the production of $^{26}$Al. Due to higher temperatures
and more convective mixing during their evolution, less $^{26}$Al
remains from the hydrostatic burning stages. However, the efficiency
of the $\nu$ process is enhanced, because of large densities in the
O/Ne shell.

The radioisotope $^{22}$Na (Figure~\ref{fig:al26s15}) is also
affected by neutral and charged current reactions.
$^{21}$Ne$(p,\gamma)$ which occurs in the O/Ne shell is enhanced by
neutral-current neutrino-induced spallation reactions and the
charged-current $^{22}$Ne$(\nu_e,e^-)$ provides a direct production
channel in the C/O layer, where $^{22}$Ne has been produced during the
He-burning phases. Compared with previous studies, the charged-current
contribution turns out to be more important for all the progenitors
studied here due to the lower neutrino energies. The total yield is
increased due to neutrinos by up to a factor of 3 (see
Figure~\ref{fig:al26na22} and Table~\ref{tab:prodall}). This effect is
very dependent on the initial conditions provided by the progenitor
and almost disappears for the 35~M$_\odot$ model. Since the different
production channels occur in spatially separated layers of the star,
the relative weight strongly depends on the position of the shell
interfaces and is therefore also sensitive to the details of the
stellar evolution.

$^{44}$Ti has been detected in supernova
remnants~\cite{Iyudin.Diehl.ea:1994,Grefenstette.Harrison.ea:2014}. It
is produced mainly in the inner ejecta in an $\alpha$-rich freeze out
of NSE~\cite{Woosley.Heger.Weaver:2002}. At high temperatures, photon-
and charged particle induced reactions dominate over any neutrino
contribution. Therefore, we find no significant effect of neutrinos on
the yield of $^{44}$Ti.  The production of $^{60}$Fe in supernovae is
discussed in detail in reference~\cite{Limongi.Chieffi:2006}, where
the neutron density reached during the shock is identified as a key
parameter for the yield. Despite the increase in the density of free
nucleons due to neutrino spallation reactions, we find no significant
modification of the $^{60}$Fe yield.

Reference~\cite{Cheoun.Ha.ea:2012} has discussed the $\nu$ process in
supernovae as a production site for the radioactive isotopes $^{92}$Nb
and $^{98}$Tc. Our calculations show, that charged-current neutrino
interactions increase the yield of $^{92}$Nb on average by a factor
of 35. The yield of $^{98}$Tc is increased by only 17\%-21\% for the
15~M$_\odot$ and the 20~M$_\odot$ progenitors. For the more massive
stars however, the enhancement goes up to a factor of 100, such that
the total yield for all of the progenitors is between $1 \times
10^{-11}$~M$_\odot$ and $6 \times 10^{-11}$~M$_\odot$. The yields for
these nuclei might even be more enhanced by contributions from the
neutrino-driven
wind~\cite{Fuller.Meyer:1995,*Hoffman.Woosley.ea:1996}.


We have performed an updated study of $\nu$ process
nucleosynthesis. Compared to previous studies, we use a full set of
neutrino-induced charged- and neutral current reactions including
spallation products for nuclei with charge numbers
$Z<76$. Additionally, we use neutrino spectra for all neutrino flavors
that are consistent with recent supernova
simulations~\cite{Huedepohl.Mueller.ea:2010,Martinez-Pinedo.Fischer.ea:2012,Martinez-Pinedo.Fischer.Huther:2014}
that predict noticeably lower average energies particularly for $\mu$
and $\tau$ (anti)neutrinos. Despite the lower average energy, we
confirm the production of $^7$Li, $^{11}$B, $^{138}$La, and $^{180}$Ta
by neutrino nucleosynthesis, albeit at slightly smaller abundances due
to the changed neutrino spectra. We find that neutrino-induced
reactions, either directly or indirectly, contribute to the production
of long-lived radioactive nuclei. The yields of $^{22}$Na and
$^{26}$Al, both prime candidates for gamma-ray astronomy, are
noticeably enhanced. As a consequence of the reduced neutrino
energies, we find that the role of charged current reactions is
enhanced with respect to previous
studies~\cite{Woosley.Hartmann.ea:1990,Heger.Kolbe.ea:2005}. The
relevant neutrino-nucleus cross-sections rely almost entirely on
theoretical calculations and are therefore accompanied by large
uncertainties. Experimental data on the relevant transitions could
help to reduce the uncertainties in order to make inferences from
observations more reliable. Furthermore, important uncertainties
remain related to the progenitor
structure~\cite{Woosley.Heger.Weaver:2002}, helium burning
rates~\cite{Austin.West.Heger:2014}, and the long term evolution of the
neutrino spectra and neutrino oscillations~\cite{Wu.Qian.ea:2015}.

\begin{acknowledgments}
  We thank Yong-Zhong Qian and Meng-Ru Wu for useful discussions.
  This work was partly supported by the Deutsche
  Forschungsgemeinschaft through contract SFB~634, the Helmholtz
  International Center for FAIR within the framework of the LOEWE
  program launched by the state of Hesse, and the Helmholtz
  Association through the Nuclear Astrophysics Virtual Institute
  (VH-VI-417). AH was supported by an Australian Research Council
  (ARC) Future Fellowship (FT120100363).
\end{acknowledgments}

%

\end{document}